\documentclass[10pt,twocolumn]{IEEEtran}

\usepackage{amssymb}
\usepackage{setspace}
\usepackage[T1]{fontenc}
\usepackage[utf8]{inputenc}
\usepackage{graphicx}
\usepackage{caption}
\usepackage{subcaption}
\usepackage{tabularx}
\usepackage{xspace}

\usepackage[linesnumbered,lined,ruled]{algorithm2e}

\usepackage[hide]{chato-notes} 

\usepackage{mathtools}
\usepackage{theorem}

\newtheorem{theorem}{Theorem}

\newcommand{\LONG}[1]{#1}
\newcommand{\SHORT}[1]{}

\title{\SHORT{On the Design of Practical Fault-Tolerant SDN Controllers}
\LONG{SMaRtLight: A Practical Fault-Tolerant SDN
Controller}}

\author{F\'{a}bio Botelho \ \ \ \ \ \ \ \ Alysson Bessani \ \ \ \ \ \ \ \ Fernando M. V. Ramos \ \ \ \ \ \ \ \ Paulo Ferreira \\
\large LaSIGE/FCUL, University of Lisbon, Portugal}

\date{}

\begin{document}
\maketitle

\begin{abstract}
The increase in the number of SDN-based deployments in production networks is triggering the need to consider fault-tolerant designs of controller architectures.
Commercial SDN controller solutions incorporate fault tolerance, but there has been little discussion in the SDN literature on the design of such systems and the tradeoffs involved.
To fill this gap, we present a by-construction design of a fault-tolerant controller, and materialize it by proposing and formalizing a practical architecture for small to medium-sized scale networks.
A central component of our particular design is a replicated shared database that stores all network state.
Contrary to the more common primary-backup approaches, the proposed design guarantees a smooth transition in case of failures and avoids the need of an additional coordination service.
Our preliminary results show that the performance of our solution fulfills the demands of the target networks.
We hope this paper to be a first step in what we consider a necessary discussion on how to build robust SDNs.
\end{abstract}

\section{Introduction}

In the past few years we have seen a steady increase in the number of SDN-based deployments in production networks.
\LONG{
Google, for example, has deployed an SDN architecture to connect its datacenters across the planet.
This production network has been in deployment for 3 years, with success, helping the company to improve operational efficiency and significantly reduce costs~\cite{jain2013b4}.
VMware's network virtualization platform~\cite{koponen2014netvirt} is another suggestive example.
This commercial solution delivers a fully functional network in software provisioned independently of the underlying network hardware, entirely based around SDN principles.
}
This initial commercial success makes imperative the need to consider scalability, availability, and resilience when building SDNs.
Fault tolerance, in particular, is an essential part of any system in production, and this property is therefore typically built-in by design. 
SDN fault tolerance covers different fault domains\,\cite{kim2012}: the data plane (switch or link failures), the control plane (failure of the switch-controller connection), and the controller itself. 
The related literature on fault tolerance in SDNs is still relatively scarce and has addressed mainly the data plane.
Kim et al's CORONET~\cite{kim2012} proposed an SDN fault-tolerant system that recovers from multiple link failures in the data plane. 
Reitblatt et al. recently proposed a new language, FatTire~\cite{Reitblatt2013}, that facilitates the design of fault-tolerant network programs.
The proposed compiler targets the in-network fast-failover mechanisms provided in recent versions of OpenFlow, and therefore the concern is again the data plane.

Failures of the control plane are of particular importance since a faulty controller can wreak havoc on the entire network.
However, to date most SDN controllers are centralized, leaving to its users the need to address the difficult challenge of guaranteeing high availability.
The first production-level SDN control platform that addresses both scalability and availability is Onix~\cite{koponen2010}.
Koponen et al.'s design resort to a physically distributed framework to achieve these goals.
Their distributed, closed-source control platform was developed for large-scale production networks, being a core component of both Google's and VMware's production systems\LONG{ mentioned above}.
Onix handles state distribution, element discovery, and failure recovery, providing a programmatic interface upon which network control planes can be built.
\note[fvramos]{Talvez se pudesse adicionar algum detalhe acerca da parte de failure recovery do Onix - pelo menos da falha de um controlador. E tambem dar desde ja uma achega de como a nossa proposta e' diferente e dar hints porque...}
\note[bessani]{Pode ser, mas nao sei como isso funciona... eles nao explicam.}

Another SDN controller worth mentioning is OpenDaylight (ODL)~\cite{Opendaylight}, an open, industry-initiated project that aims to devise a distributed network controller framework.
The work of incorporating fault tolerance primitives in the ODL controller is undergoing, with a preliminary working cluster based on the use of Infinispan datastore~\cite{Infinispan} to tolerate faults.
However, its unsatisfactory performance have led to a rethinking of the solution.
Currently, ODL is still far from settling on a specific design and implementation in this respect.

\note[bessani]{Acima, talvez fosse bom ter a referencia precisa para thread onde discutem que a solucao do infinispan nao funciona e que tem de pensar em outras coisas.}

This paper intends to be a first discussion on the design and implementation of practical fault-tolerant SDN architectures and the tradeoffs involved thereof.
Onix provided a first insight into this problem, but being a complex platform targeting large-scale environments it had to deal with issues that ranged from availability and scalability to generality.
The proposal we make here is more modest, but simultaneously more focused and detailed on the fault tolerance aspect.
We intend with this paper to cover ground on fault-tolerant protocols and mechanisms that have received little attention, to the best of our knowledge, in the context of SDN.
We hope the discussion ensued offers guidelines to the design of controllers (such as ODL) that intend to incorporate fault tolerance mechanisms into their architecture.

To this aim, we describe a by-construction design of a simple, fault-tolerant controller architecture for small to medium-sized networks.
In this architecture the network is managed by a single controller, and other controllers are used as backups that can take over its role in case the primary fails.
In case a failure occurs, and to ensure a smooth transition to a new primary, in this particular instance of a fault-tolerant architecture we make the controller store the network- and application-related state in a shared data store~\cite{Bot13}.
This data store is implemented as a Replicated State Machine (RSM)~\cite{Bes14,Lam98,Sch90}, and is therefore also fault-tolerant.
Importantly, the protocol does not require strong synchrony assumptions and ensures safety in any situation\SHORT{.}\LONG{, but allows bounded and fast primary replacement when the network is stable.}

In previous work we have shown that such data store meets the performance demands of small to medium SDNs~\cite{Bot13}.
We further demonstrated that the state-of-the-art distributed systems techniques (e.g., \cite{Bes13,Hun10}) we used to optimize throughput has resulted in a significant performance improvement over the values reported in the Onix paper for its consistent data store.
To further increase the performance and scalability of our solution, in this work we incorporate a cache in the controllers to avoid accessing the shared data store on reads.

In summary, the main contribution of this paper is the discussion of a pragmatic design for a centralized, fault-tolerant, SDN controller, and the precise description of the required assumptions and properties for such design to hold.
Put in order words, we discuss the required building blocks to cost-effectively transform a centralized SDN controller into a fault-tolerant architecture.

\section{FT controller architecture}



In this section we present a by-construction description of a fault-tolerant controller.
We start by considering an SDN in which a single controller is used to manage all network switches.
This is still the most common setting and its feasibility for medium-sized networks has been demonstrated before~\cite{Tootoonchian:2012:CPS:2228283.2228297}.
To ensure the network is controlled despite faults in this controller, we leverage the multi-controller capabilities of OpenFlow 1.3~\cite{ONF13} and deploy the same applications in several controller replicas. 
All switches therefore establish a connection with all controllers.
 
With several controllers deployed, we use a single one to manage the network (the \emph{primary}) whereas the others are used as backups that can take over the role of primary in case of its failure.
To enable this fault-tolerant solution it is fundamental to have algorithms for fault detection and leader election~\cite{Hun10}.
With this solution we have a fault-tolerant control plane in which a crashed controller will be replaced by one of its replicas.

An important limitation of this solution is the fact that the new primary will have an empty state after it starts managing the network.
\note[fvramos]{Aqui o prob nao sera o dropout dos flows, pq eles estao instalados nos switches, pelos menos durante algum tempo. O problema serao as inconsistencias entre o que o controlador sabe da rede e o que de facto se passa.}
One way to solve this issue is to make the controller replicas store the network- and application-related state (the Network Information Base or NIB) in a shared data store~\cite{Bot13}.
In this scenario, the primary controller always updates the NIB before modifying the state of the network.
This ensures a smooth transition to a new primary. 
When the old primary fails, the new primary takes over its role and its first action is to read the current state from the data store.
As the network state in the data store is always up-to-date, the new primary will have a consistent view of the network from the outset.

The final problem that needs to be addressed is how to implement such data store, to ensure that it is not a single point of failure and that it provides sufficient performance to be used in real networks.
A possible solution relies on the use of the Paxos algorithm~\cite{Lam98} for implementing the data store as a Replicated State Machine (RSM)~\cite{Sch90}.
The use of a RSM gives us the guarantees that a data store update will be reflected in successive reads by a different process, ensuring thus that no update performed by the primary on a data store will be lost after a failure (assuring the smooth transition).
As mentioned before, we have shown that such data store can be efficiently implemented~\cite{Bes13,Hun10} to the point of satisfying the performance demands of small to medium-scale SDNs~\cite{Bot13}.

\begin{figure}[!htp]
\begin{center}
\includegraphics[width=\columnwidth]{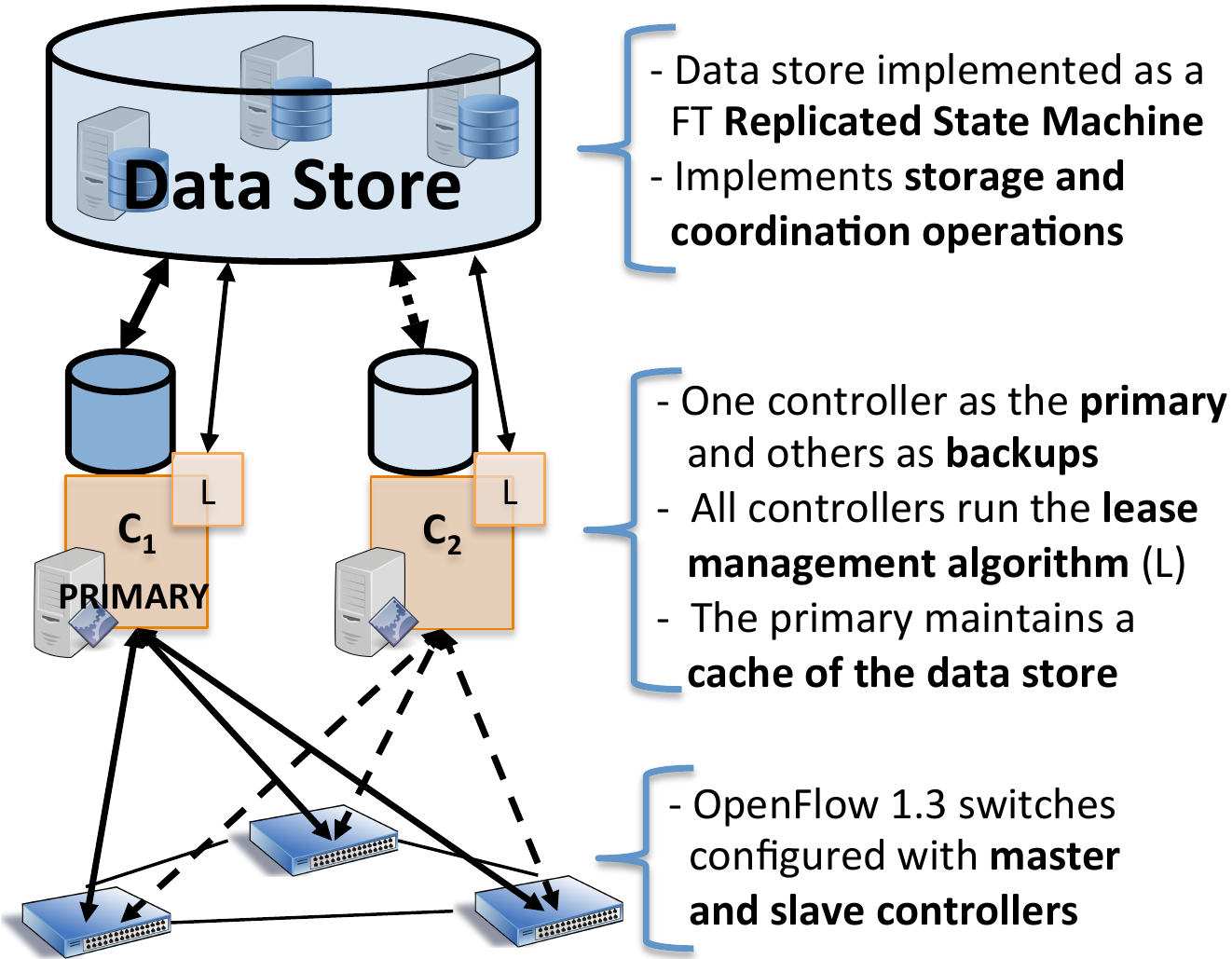}
\end{center}
\caption{SMaRtLight architecture.}
\label{fig:architecture}
\end{figure}

\noindent \textbf{SMaRtLight architecture.}
In order to materialize the techniques here described, we have designed and implemented a fault-tolerant controller -- SMaRtLight.
The architecture of SMaRtLight is presented in Figure~\ref{fig:architecture}.
We leverage on a RSM shared data store~\cite{Bot13} to avoid the empty state problem explained before. 
The architecture includes two additional aspects that we have not discussed thus far.
First, we integrate the fault detection and leader election algorithms using the data store to avoid the use of an additional coordination service.
An alternative solution would be to implement the data store in the coordination service.
However, using a pre-defined service as a data store constrains control applications to use the interface provided by this service.
In our experience, this constraint may have a cost in performance.
With our solution we have more freedom to implement the operations that best fit the application.
Second, the controllers keep a local cache to avoid accessing the shared data store in the case of read operations.
This cache does not require synchronization since there is only one active primary accessing the data store at any one time.

\section{System model and assumptions}
\label{sec:model}

In order to precisely define the control plane fault tolerance capabilities of our system we first formalize the system model and service properties.

We consider a system composed of three sets of processes.
There is a set $\mathcal{S}$ with an undefined number of switches, a set $\mathcal{C}$ with $|\mathcal{C}|$ controllers and a set $\mathcal{D}$ with $|\mathcal{D}|$ data servers.

We assume that processes in $\mathcal{S}$ can communicate with processes in $\mathcal{C}$, but not with processes in $\mathcal{D}$.
Furthermore, processes in $\mathcal{C}$ can communicate with processes in $\mathcal{S}$ and in $\mathcal{D}$, which themselves can communicate with processes in $\mathcal{C}$ and in $\mathcal{D}$.
\note[fvramos]{Aqui quer-se dizer que os processos em D comunicam com os processos em D, certo? Fiquei na duvida, talvez se possa clarificar.}
This communication is done through \emph{fair channels}, i.e., channels that can drop messages but that will eventually deliver a message if it is retransmitted a sufficient number of times.
We say a process $p$ is \emph{connected} with a process $q$ if a message sent by $p$ is answered (or acknowledged) by $q$ within an application-defined time interval.
In practice, the notion of ``connected'' as defined above aims to capture the processes that can establish a TCP connection.

We assume a partially synchronous (communication and processors) system model~\cite{Dwo88} in which there are no time bounds in either the communication or the local processing.
However, at a certain (unknown) instant $GST$ (Global Stabilization Time) the system becomes synchronous, starting to respect some (unknown) processing and communication time bounds.
In practice, the system does not need to be synchronous forever, but only during the execution of the protocol of interest.

This model corresponds to the weakest synchrony assumptions required for implementing a consistent and responsive fault-tolerant data repository for the system~\cite{Dwo88}.
Furthermore, it is a nice approximation for best-effort networks, which usually present a stable and predictable behavior, but sometimes can be subject to perturbations that make its behavior unpredictable.

Finally, we assume all processes are equipped with local (non-synchronized) clocks with a drift bounded by $\delta$.
This clock is used for running periodic tasks.

All processes in the system can crash and later recover.
We precisely define the notion of \emph{correct} process in the following way:

\begin{itemize}
\item A \emph{switch is correct} if it is not crashed and is connected with all correct processes in $\mathcal{C}$;
\item A \emph{controller is correct} if it is not crashed and is connected with all correct data servers.
\item A \emph{data server is correct} if it is not crashed or recovering (i.e., running a recovering protocol~\cite{Bes13}), and is connected with all correct data servers.
\end{itemize}

A process that is not correct is said to be \emph{faulty}.
Our definition of correct processes explicitly assumes that a correct process is able to reliably communicate with other elements of the system.
In this way, a process that is not connected with others is considered faulty according to our definition.

In this paper we do not consider data plane fault tolerance, and assume this to be dealt by SDN applications~\cite{kim2012,Reitblatt2013} (e.g., through the identification of faults and reconfiguration of routes).
Consequently, we assume that the network ``works'' as long as there are at most $f_s$ faulty switches in the system.
Furthermore, we consider that at any moment during system operation there are at most $f_c < |\mathcal{C}|$ faulty controllers (at least one must be correct) and $f_d < \frac{|\mathcal{D}|}{2}$ faulty data servers.
This last threshold defines the need of a majority of correct data servers.
Together with the partially synchronous system model, this is a requirement for ensuring safety and liveness of the replicated data store~\cite{Lam98}.

\section{Design and implementation}

In the following sections we detail how the switch, controller and data store functionality are integrated in the SMaRtLight architecture.

\subsection{Switches}

We consider that switches can be attached to multiple controllers, as defined in OpenFlow 1.3~\cite{ONF13}.
The basic idea is for each switch to maintain an open connection to every controller.
Initially, all switches consider all controllers as having the role \texttt{OFPCR\_ROLE\_EQUAL}.
When the controller replicas elect the primary (see next section), this controller sets its role in each switch to \texttt{OFPCR\_ROLE\_MASTER}.
This causes the switches to change the role of all other controllers to \texttt{OFPCR\_ROLE\_SLAVE}.
With this configuration, the switches accept commands and send event notifications exclusively to the primary controller.
In case of its failure, the backup controllers elect a new primary. 


\subsection{Controller replicas}

The controller replicas are common SDN controllers (e.g.,~\cite{Floodlight,Gude:2008jd}) deployed with the same applications in different servers.
To ensure their correct operation despite failures and asynchrony, the controllers run a small coordination module that enforces two properties:

\begin{itemize}
\item \textbf{Safety:} At any point in the execution, there is at most one primary controller $p \in \mathcal{C}$.
\item \textbf{Liveness:} Eventually some correct $p \in \mathcal{C}$ will become a primary.
\end{itemize}

This module, represented by the L box in Figure~\ref{fig:architecture}, is deployed in the controller as an application that does not interact with the switches, only with the data store. 

The module runs an algorithm that periodically calls an $acquireLease(id,L)$ operation on the data store.
This operation receives the caller $id$ and a requested lease time $L$ as parameters and returns the id of the current primary.
Moreover, if there is no primary defined or the lease is expired, the operation returns the invoker id, making it the primary for the requested lease time.
Finally, if the invoker is the lease owner, the operation returns the caller id and renews the lease for  $L$ time units.
The implementation of this operation in a replicated state machine is described in the next section.
Making use of this operation, the coordination module implements both leader election and fault detection by running Algorithm~\ref{alg:leaderelection}.

\begin{algorithm}[!htb]
{\small
\caption{Lease management on a controller $c$.}
\label{alg:leaderelection}
\DontPrintSemicolon
\textbf{Initialization of global variable at startup:}\\
\SetAlgoVlined
\Begin{
	$primary \leftarrow null$\;
	$my\_lease \leftarrow 0$\;
}
\textbf{Coordination task}\\
\Begin{
	\Repeat{shutdown}{
		$start \leftarrow clock()$
		$curr\_primary \leftarrow datastore.acquireLease(c,L)$\;
		\If{$curr\_primary = c$}{
			\If{$primary \neq c$}{
				\ForEach{$s \in \mathcal{S}$}{
					send $\langle c,\texttt{OFPCR\_ROLE\_MASTER} \rangle$ to $s$\;
				}
			}
			$my\_lease \leftarrow start + L$\;
			\lIf{$my\_lease < clock()$}{
				$L \leftarrow 2 \times L$\;
			}
		}
		$primary \leftarrow curr\_primary$\;
		$sleep(\Delta-(clock()-start))$\;
	}
}
\textbf{predicate} $iAmPrimary() \equiv my\_lease > clock()$
}
\end{algorithm}

Periodically, every $\Delta < L$ time units,\footnote{During periods of synchrony, the fault detection time is bounded by $L+\Delta$.} each replica (primary or backup) invokes $acquireLease()$.
In the backup, if the returned id corresponds to itself, the replica is the new leader and must change its role to \texttt{OFPCR\_ROLE\_MASTER} in all switches (lines 10-12).
Otherwise, nothing happens.
In the primary, if the returned id corresponds to itself, the lease was renewed, and $my\_lease$ is updated (line 13).
Otherwise, the controller lost the primary role (due to its lease being expired) and should not interact with the switches or the data store.

\begin{figure}[!htp]
\begin{center}
\includegraphics[width=0.9\columnwidth]{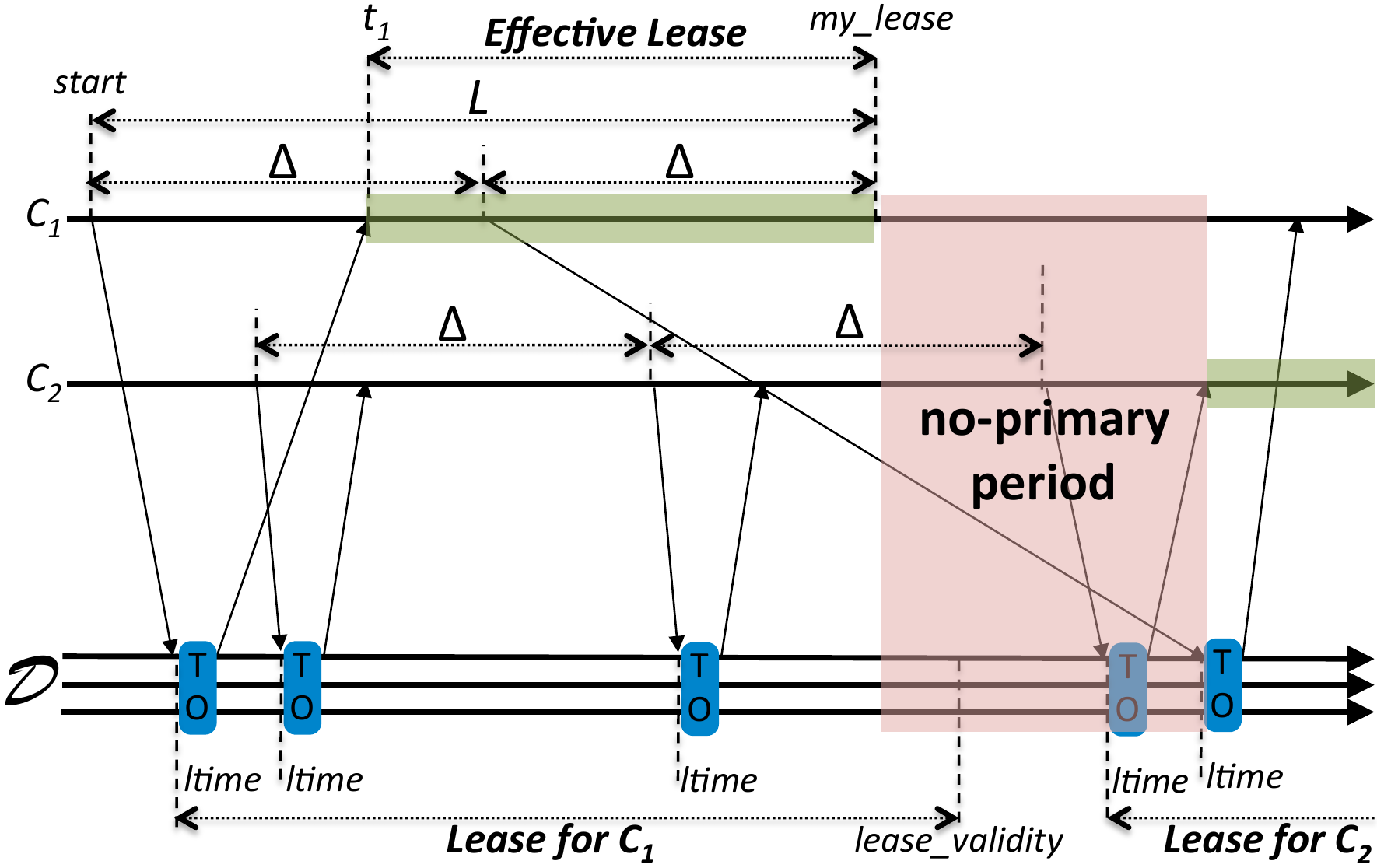}
\end{center}
\caption{Lease and effective lease in SMaRtLight.}
\label{fig:lease}
\end{figure}

Besides running this algorithm, the controller needs to be modified to only interact with switches or the data store if the predicate $iAmPrimary()$, which tests if the controller has a non-expired lease, is true.
Notice that  $my\_lease$ is defined based on $start$ -- the instant right before invoking $acquireLease()$ on the data store.
This ensures the \emph{effective lease} time on the replica to be smaller than the \emph{lease granted} by the data store, as illustrated in Figure~\ref{fig:lease}.
\emph{Safety} is ensured since at any point of the execution (even in periods of asynchrony), there will be \emph{at most} one primary controller.

In order to ensure \emph{liveness}, i.e., that eventually some controller will be elected as primary, a controller increases its requested lease time $L$ every time it receives an expired lease (line 14).
This happens in periods in which accessing the data store takes more than $L$ time units, thus not being safe to use the lease even if it is granted to the invoker.

\SHORT{In the extended version of this paper~\cite{Bot14-ext} we present proofs for these two properties.}
\LONG{In the appendix we present proofs for these properties.}

\subsection{Data Store}
\label{sec:datastore}

In order to implement a fault-tolerant data store with good performance we employ a Paxos-like protocol for building replicated state machines~\cite{Lam98}.
In a nutshell, in a replicated state machine all replicas start on the same state and run the same deterministic code~\cite{Sch90}.
The replication protocol delivers the same client requests in the same order to all correct replicas.

For the purpose of this work, we consider the replication protocol as a black box and specify the deterministic code that each replica must run.
This code corresponds to the $acquireLease()$ operation and to all operations required by the applications running in the controller (typically, key-value store operations for manipulating the shared state).

An important feature required is that a timestamp $ltime$ should be delivered together with the $acquireLease()$ requests (also illustrated in Figure~\ref{fig:lease}).
This timestamp is the real-time clock value of the Paxos leader~\cite{Lam98}, which is read just after receiving the client (i.e., controller) request, as is common in similar replicated systems~\cite{Bes14,Cas02} to ensure deterministic timestamping.
All replicas use this $ltime$ instead of their clock value to validate and calculate the granted lease times, which ensures that they will reach the same decision regarding the lease validity.
The code for this operation in the replicas is described in Algorithm~\ref{alg:acquirelease}.

\begin{algorithm}[!ht]
{
\footnotesize
\small
\caption{$acquireLease()$ on a data server.}
\label{alg:acquirelease}
\DontPrintSemicolon
\textbf{initialization of global variable at startup:}\\
\SetAlgoVlined
\Begin{
	$primary \leftarrow null$\;
	$lease\_validity \leftarrow 0$\;
}
\textbf{when} $\langle \langle \mathsf{ACQUIRELEASE},id,L \rangle, ltime \rangle$ is delivered\\
\Begin{
	\If{$lease\_validity > ltime$}{ 
		\lIf{$primary = id$}{
			$lease\_validity \leftarrow ltime + L$\; 
		}
	}\Else{
		$primary \leftarrow id$\; 
		$lease\_validity \leftarrow ltime + L$\; 
	}
	\Return $primary$\;
}
}
\end{algorithm}

Besides the $acquireLease()$ coordination primitive, the data store needs to implement the shared storage for the controllers.
This storage is based on a key-value store interface that supports simple operations such as put, get, remove, and list, and on the use of a main-memory cache at the client (controller) side.
\LONG{Figure~\ref{sec:datastore} shows the interaction between the cache and the data store.

\begin{figure}[!htp]
\begin{center}
\includegraphics[width=0.9\columnwidth]{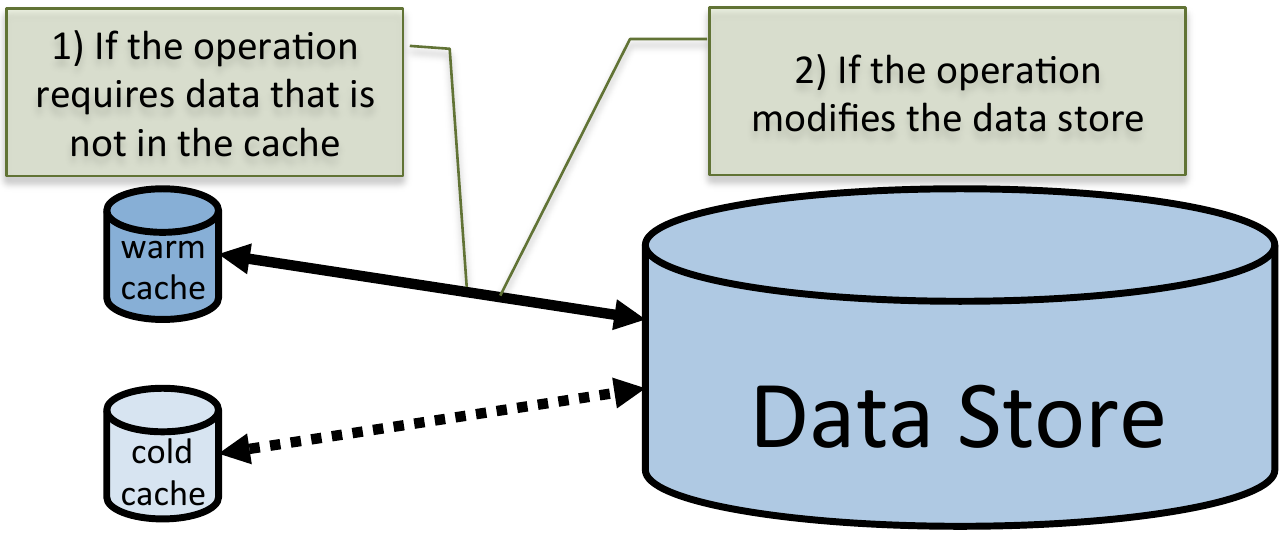}
\end{center}
\caption{Cache and data store interaction.}
\label{fig:datastore}
\end{figure}
}

Only the primary controller interacts with the data store, reading from or writing to the application-required tables.
Simultaneously, it also updates its cache.
A backup controller has this cache empty (cold) until becoming the primary.
When a control application needs to read some information from the data store it first tries on the cache and only when a cache miss occurs is the data retrieved from the data store.
Write operations (i.e., operations that modify the data store state) complete only after the cache and the application table(s) in the data store are updated.

We antecipate our architecture to make efficient use of the cache for two main reasons.
First, we expect the size of a NIB not to be significant when compared with the amount of main-memory available on controller machines.
Therefore, in most cases, the primary will have the whole copy of the data store in its main-memory, absorbing all reads locally.
Second, the fact that we have a single process (the primary) updating the cache allows its extensive use without requiring complex and low-performant cache invalidation protocols.

\subsection{Implementation}

We implemented a prototype of this architecture by integrating the Floodlight~\cite{Floodlight} centralized controller with a simple extensible key-value data store built over BFT-SMaRt~\cite{Bes14}, a state-of-the-art state machine replication library.
When configured for crash-fault tolerance, this library runs a protocol similar to Paxos.

The Floodlight controller was extended in five ways.
First, we implemented the lease management as an application of the system.
Second, our prototype maintains a one-to-one mapping between switches and data store connections to enhance the parallelism of the controller.
Third, we modified the controller-switch communication to allow interaction only if the predicate $iAmPrimary()$ is true (see Algorithm~\ref{alg:leaderelection}). 
Fourth, since Floodlight only supports OpenFlow 1.0, we extended the controller to be able to send (OF 1.3) change role messages.
Finally, we re-implemented the data store interface for supporting our data store model with caching.
Each controller maintains tables in its local cache to keep the most frequently accessed data.

\section{Preliminary Evaluation}

Our preliminary evaluation investigates two questions: \emph{How does the introduction of a fault tolerant data store affect the performance of a centralized controller? What is the performance impact of controller faults?} 

We use CBench~\cite{Tootoonchian:2012:CPS:2228283.2228297} to simulate a number of switches sending \texttt{packet-in} messages to the controller.
Since this tool does not assume any application to be running on the controller and our main bottleneck is expected to be data store access, we developed a dummy application that processes requests locally with a probability $P$ (using the cache), accessing the data store with probability $1-P$.
Every time the data store is accessed a payload of 44 bytes is atomically written and read.

\note[fvramos]{Qual o rationale para os 44 bytes?}

Different values of $P$ can be used to represented different application behaviors.
For example, a topology manager is accessed for each new flow but only updates the data store when the topology changes, so it would extensively use its cache ($P \geq 0.9$).
A load balancer, on the other hand, will have to ensure the data store always contains up-to-date information on the last server to whom flows were assigned (hence, $P=0$).



We used two controllers and three data store replicas in all experiments, tolerating a single fault in each of these layers.
Each of these replicas runs on quad-core 2.27 GHz Intel Xeon E5520 machines with 32 GB of RAM, and are interconnected with a gigabit Ethernet. 
The software environment is composed of Ubuntu 12.04.2 LTS and Java(TM) SE Runtime Environment (build 1.7.0 07-b10) 64 bits.
We set the algorithm parameters to $\Delta=0.5$ s and $L=1$ s.

\begin{figure*}[ht]
\centering
\begin{subfigure}[b]{.45\textwidth}
  \centering
  \includegraphics[width=\textwidth]{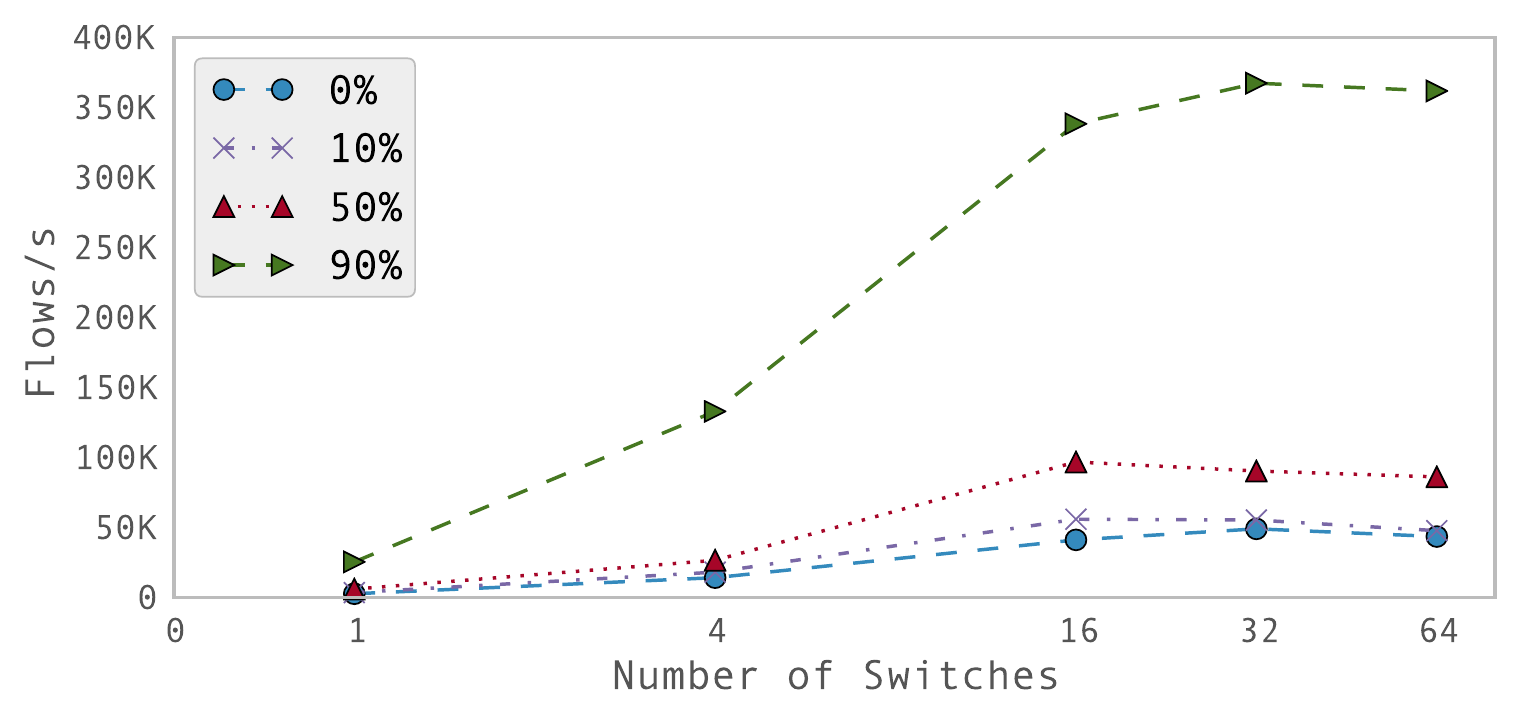}
  \caption{Throughput with different cache effectiveness.}
  \label{fig:performance}
\end{subfigure}%
\hspace{4mm}
\begin{subfigure}[b]{.45\textwidth}
  \centering
  \includegraphics[width=\textwidth]{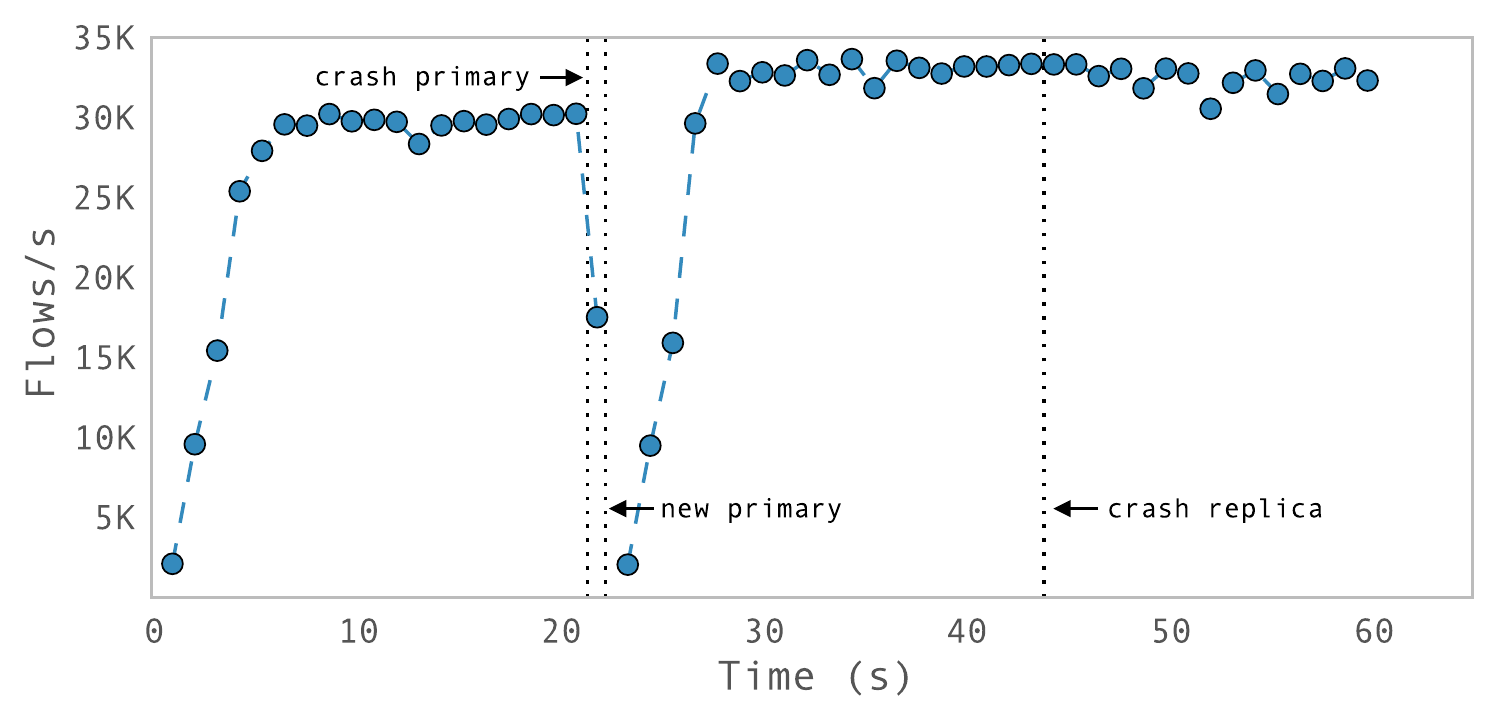}
  \caption{Throughput with faults.}
  \label{fig:fault}
\end{subfigure}
\label{fig:results}
\caption{Raw performance and fault injection evaluation. }
\end{figure*}

\noindent \textbf{Raw performance.}
We configured CBench to run 10 experiments of 10 seconds each, with a varying number of switches (1-64) and a fixed number of simulated hosts (1000).  
To achieve maximum performance, we deployed CBench in the same machine as the primary controller, since the generated traffic exceeded the NIC bandwidth. 

Figure \ref{fig:performance} shows the throughput obtained for a varying number of switches. 
We show results considering $P = 0$, $0.1$, $0.5$ and $0.9$. 
The figure shows that with 90\% of operations absorbed by the cache,
the system can process up to 367K Flows/sec.
For write-heavy applications the achieved throughput decreases to 96K Flows/sec (50\%) and 55K Flows/sec (10\%).



Even considering this preliminary version of SMaRtLight, we found the results promising: SDN applications can be made fault-tolerant while still securing a processing rate between 100K Flows/s to 350K Flows/s, as long as they provide cache hits rates above 50\%.

These values are still behind what can be obtained with multi-threaded, centralized controllers (according to our measurements, Floodlight can process up to 2.5M flows/s).
Anyway, there is still room for improvement in our implementation.
For example, in the most demanding scenarios, we imposed up to 48K (44 byte) updates/sec to the data store, which corresponds to less than 70\% of the expected capacity of BFT-SMaRt for such workload~\cite{Bes14}.

More importantly, in SMaRtLight each  update is performed on the data store to guarantee that it survives the failures of either the controller or a data store replica. 
To better illustrate the impact of this difference, we measured the throughput of a Learning Switch application modified to (synchronously) append NIB updates to a disk log (for recovery after a failure). 
This experience showed a performance of 200 flows/s for 32 switches, which is $250\times$ worse than what is provided by our data store without cache. 

\noindent \textbf{Effect of faults.}
In order to test the fault tolerance capabilities of SMaRtLight we modified CBench to simulate the behavior of a switch connected to multiple controllers.
Initially the switches generate load to the primary controller, but change to the backup after receiving a role-change request from this controller.

Figure \ref{fig:fault} shows the observed throughput of a setup with 10 switches and $P=0$ (no cache) during an execution in which the primary controller and one data store replica fail.
Although the backup replica takes over the role of primary in less than 1 second, it takes around 4 seconds for the system to return to its normal throughput.
This corresponds to the time for the new primary to inform the 10 switches about the change and for them to start generating load.\footnote{This is also due to Java's JIT compiler that makes the new primary controller slower until it warms-up (this can also be seen at the beginning of the experiment).}
Notice also that the crash of a data store replica does not affect the performance of the system.


\section{Discussion}
\label{sec:discussion}

The general approach to building fault tolerant systems is redundancy. 
To enable a Software-Defined Network to tolerate the loss of a controller, we need to add extra equipment: additional controller replicas.
There are  two fundamental approaches for replication: primary-backup (or passive) replication~\cite{alsberg1976} and state machine (or active) replication~\cite{lamport1978}.
In the primary-backup replication model there is one primary replica that executes all operations issued by clients and, periodically, pushes state updates to the backup replicas.
These replicas keep monitoring the primary, to ensure one of them takes over in case it fails.
In the state machine replication model clients issue commands to all replicas that execute them in a coordinated, deterministic way.
In the design of our fault-tolerant controller we opted for the latter approach because passive replication techniques can suffer from poor performance (relative to active replication) in the presence of failures.
Passive replication suffers from high reconfiguration costs when the primary fails.
The main drawback generally attributed to active replication is the requirement that operations on the replicas be deterministic.

In order to ensure that the network is unaffected by a controller failure, there are two important issues that need to be addressed.
First, that every controller, primary or backup, always have the same view of the network.
Second, that in the event of a controller failure all other replicas agree on which to replace it.
The first issue can be solved by implementing a shared datastore, as we did.
Although we chose to develop a datastore on top of a state machine replication library, an alternative datastore could have been used (e.g., Cassandra~\cite{Lakshman2010}).
\note[pferreira]{justificar a escolha do smart? dar exemplos de alternativas (ex: Cassandra)? Falar de consistency?}
Regarding the second issue, a coordination service such as ZooKeeper~\cite{Hun10} could have been used to select the next controller to act as primary.
However, we have decided to take advantage of BFT-SMaRt consistency semantics and therefore embedded the coordination logic inside the data store itself. 

Another alternative would be to use a coordination service like ZooKeeper as the data store, instead of implementing the coordination primitive ($acquireLease()$) in a custom data store, as we did.
The widespread adoption of that service and its proved robustness are valid arguments to opt for such approach.
We opted for having such service implemented on top of our replication library because we found it simpler\footnote{In part, due to our knowledgeability of BFT-SMaRt.} to improve system performance by implementing specific operations that best fit the network application.
For example, to implement a round-robin load balancer one first needs to read the server that will respond to the next request and then update this information to define the next server.
This corresponds to two operations in  Zookeeper but can be easily implemented with a simple ``read and increment'' operation in our data store.

\section*{Acknowledgements}
Thanks to Diego Kreutz, Regivaldo Costa and Paulo Verissimo for the comments that helped improve the paper.
This work was supported by the EC FP7 through project BiobankCloud (ICT-317871) and by FCT through the LaSIGE Strategic Project (PEst-OE/EEI/UI0408/ 2014).

{\footnotesize
\bibliographystyle{abbrv}
\bibliography{ewsdn14}
}

\LONG{
\appendix

\section{Correctness Proofs}
{\small
\emph{``That taught me a lesson, that I should not write a concurrent algorithm without having a proof of its correctness.''} -- Leslie Lamport (about his first attempt to solve mutual exclusion)
}

In the following we present the proofs of safety and liveness of SMaRtLight algorithms.
To prove the safety we use a function $\mathsf{C}()$ that maps the local clock values in the processes to the (global) real-time clock value.
Note that the processes do not have access to this function, this is only a theoretical device to reason about the correctness of the system.

\begin{theorem}[Safety]\label{t:safety}
At any point in the execution, there is at most one primary controller $p \in \mathcal{C}$.
\end{theorem}

\noindent \emph{Proof:}
We will prove this theorem by contradiction.
Assume that there exists a global real-time instant $t$ in which two controllers $c_1, c_2 \in \mathcal{C}$ both consider themselves as the primary.
Let $t_1$ (resp. $t_2$) be the local time $acquireLease()$ returned to $c_1$ (resp. $c_2$) and 
the $my\_lease$ variable of $c_1$ and $c_2$ on instant $t$ be represented by $v_1 = t_{1start}+L$  in $c_1$ and $v_2 = t_{2start}+L$ in $c_2$, respectively.
Within these definitions, the existence of $t$ requires that either $\mathsf{C}(t_2) \leq  \mathsf{C}(v_1)$ or  $\mathsf{C}(t_1) \leq \mathsf{C}(v_2)$, as illustrated in Figure~\ref{fig:proof-illustration}.

\begin{figure}[!htp]
\begin{center}
\includegraphics[width=\columnwidth]{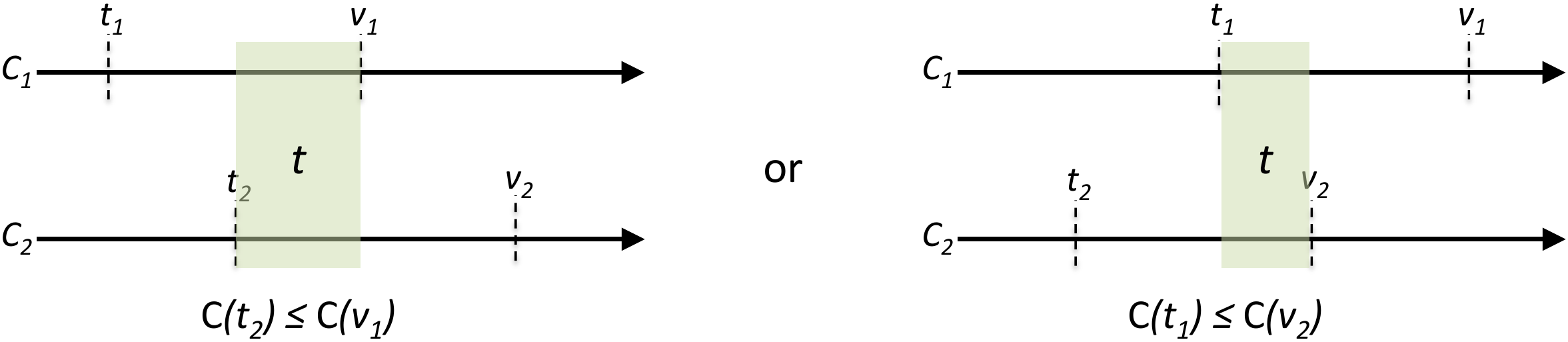}
\end{center}
\caption{Safety proof illustration.}
\label{fig:proof-illustration}
\end{figure}

Since the data store process requests in total order, it either executed Algorithm 2 for the $c_1$ request (with local time $ltime_1$) and later for the $c_2$ request (with local time $ltime_2$), or vice-versa.
Assume, without loss of generality, that it first executed this operation for $c_1$ and, during the operation processing, it sets $lease\_validity = ltime_1 + L$ and $primary = c_1$ and returns $c_1$, which is received by this controller at local time $t_1$.
Intuitively, this corresponds to the left case of Figure~\ref{fig:proof-illustration}.

Notice however that since the invocation of $acquire$ $Lease()$ precedes its reception by the RSM leader (which defines $ltime_i$) and the processing of a request precedes its reception by $c_i$.
Consequently, we have:

\begin{align}
\mathsf{C}(t_{1start}) < \mathsf{C}(ltime_1) < \mathsf{C}(t_1) \label{eq:c1}\\
\mathsf{C}(t_{2start}) < \mathsf{C}(ltime_2) < \mathsf{C}(t_2) \label{eq:c2}
\end{align}

Controller $c_2$ will only receive its id when accessing the data store, if the latter observes $ltime_2 > lease\_validity = ltime_1 + L$, it sets update its lease validity and changes $primary$ to $c_2$, returning also this value.
If this condition holds, we have:

\begin{align}\label{eq:ds}
\mathsf{C}(lease\_validity) < \mathsf{C}(ltime_2)  \xRightarrow{implies} \nonumber \\
\mathsf{C}(ltime_1 + L) < \mathsf{C}(ltime_2)  \xRightarrow{implies} \nonumber \\
\mathsf{C}(ltime_1) + L < \mathsf{C}(ltime_2)
\end{align}

Using the results from Equations \ref{eq:c1} and \ref{eq:c2} together with Equation \ref{eq:ds} we can reach:

\begin{align}\label{eq:result-c1first}
\mathsf{C}(t_{1start}) +L < \mathsf{C}(t_2)  \xRightarrow{implies} \nonumber \\
\mathsf{C}(t_{1start} + L) < \mathsf{C}(t_2)  \xRightarrow{implies} \nonumber \\
\mathsf{C}(v_1) < \mathsf{C}(t_2)
\end{align}

Equation~\ref{eq:result-c1first} directly contradicts one of the conditions for the existence of $t$ (left of Figure~\ref{fig:proof-illustration}).
To prove that the other condition is also impossible we have to use the same approach, assuming that the $c_2$ request for lease is processed before the $c_1$ request by the data store.
\hfill $\blacksquare$

\begin{theorem}[Liveness]\label{t:liveness}
Eventually some correct $p \in \mathcal{C}$ will become a primary.
\end{theorem}

\noindent \emph{Proof (sketch):}
To prove this we have to show that eventually some controller will (1) receive its id as a reply from the data store and (2) the time of reception will still allow some effective lease (see Figure~\ref{fig:lease}).

Consider that (a) each controller keeps trying to acquire the lease (loop of Algorithm~\ref{alg:leaderelection}) and (b) after $lease$ $\_validity$ from the last successful lease granting, the data store will grant the lease to the first controller who asks for it (Algorithm~\ref{alg:acquirelease}).
(a) and (b) together ensure (1), i.e., that some controller will eventually be granted a lease.

Now we have to show that eventually the controller will receive the lease grant confirmation before this lease expires.
For proving this, we have to use our assumption of partially synchronous communication and process system model (see \S\ref{sec:model}).
During periods of asynchrony, the system can be very slow and a replica can receive an already expired lease.
In this case, the replica doubles its next requested lease time (line 14 of Algorithm~\ref{alg:leaderelection}).
When the system becomes synchronous, there will be unknown bounds on communication and processing, which will reflect in a bounded response time $B$ for $acquireLease()$.
If $B > L$, the replicas will keep increasing their $L$ until one of them is able to acquire the lease.
\hfill $\blacksquare$}
\end{document}